\documentstyle[epsf,epsfig]{elsart}  
\begin{document}
\newcommand{\p}{\partial}
\newcommand{\ls}{\left(}
\newcommand{\rs}{\right)}
\newcommand{\beq}{\begin{equation}}
\newcommand{\eeq}{\end{equation}}
\newcommand{\beqa}{\begin{eqnarray}}
\newcommand{\eeqa}{\end{eqnarray}}
%%%%%%%%%%%%%%%%%%%%%%%%%%%%%%%%%%%%%%%%%%%%%%%%%%%%%%%%%%%%%%%%%%%%%%%%%
%                                                                       %
%   BEGIN OF DOCUMENT                                                   %
%                                                                       %
%%%%%%%%%%%%%%%%%%%%%%%%%%%%%%%%%%%%%%%%%%%%%%%%%%%%%%%%%%%%%%%%%%%%%%%%%
\begin{frontmatter}
\title{Temperature and thermodynamic instabilities in 
heavy ion collisions} 
\author[tuebingen]{C. Fuchs},
\author[muenchen]{P. Essler},
\author[muenchen]{T. Gaitanos},
\author[muenchen]{H.~H. Wolter}
\address[tuebingen]{Institut f\"ur Theoretische Physik der 
Universit\"at T\"ubingen, D-72076 T\"ubingen, Germany}
\address[muenchen]{Sektion Physik, Universit\"at M\"unchen, 
D-85748 Garching, Germany}  
%************************************************************************
\begin{abstract}
We investigate  thermodynamic properties and instability conditions in
intermediate 
energy heavy ion reactions. We define locally thermodynamic 
variables, i.e. density, pressure and temperature, directly from the 
phase space distribution of a relativistic transport calculation. 
In particular, temperatures are determined by a fit to two covariant 
hot Fermi distributions thus taking into account possible 
anisotropic momentum configurations. 
We define instability independent from the nuclear matter spinodal by the
criterion that the effective compressibility becomes negative.  The method is
applied to a semi--central Au on Au reaction at 600~MeV/nucleon. We
investigate in particular the center of the participant and the spectator
matter. In the latter we find a clear indication of instability with
conditions of density and temperature that are consistent with experimental
determinations.
\end{abstract}  
%************************************************************************
\end{frontmatter}
%%%%%%%%%%%%%%%%%%%%%%%%%%%%%%%%%%%%%%%%%%%%%%%%%%%%%%%%%%%%%%%%%%%%%%%%%
%                                                                       %
%   BEGIN OF TEXT                                                       %
%                                                                       %
%%%%%%%%%%%%%%%%%%%%%%%%%%%%%%%%%%%%%%%%%%%%%%%%%%%%%%%%%%%%%%%%%%%%%%%%%
\section{Introduction}
%%%%%%%%%%%%%%%%%%%%%%%%%%%%%%%%%%%%%%%%%%%%%%%%%%%%%%%%%%%%%%%%%%%%%%%%
One of the challenges in the investigation of heavy ion collisions is the
understanding of the multi--fragmentation process which is observed in the
final stages of such reactions. The mass spectra are observed to follow a
power law which leads to the concept of a first--order liquid--gas phase
transition or even of  second--order critical behavior. Recently a first
order phase transition seems to have been found by the ALADIN collaboration
\cite{poch95,traut97} from spectator fragments in 
Au + Au collisions at 600~A.MeV,
but this interpretation is still under much debate, in particular with
respect to the correct thermometers to be used \cite{traut97}.

Equally the theoretical description of the fragmentation process is widely
debated. The models can be roughly divided in two classes, 
dynamical models based on transport equations 
\cite{ag90,rr90,abe96,colonna,bonasera,puri95} and statistical models based on
the assumption of  local thermal equilibrium \cite{bon85,gross}. 
Transport models as BUU, however, are mean field models which describe the
evolution of the one--body phase space density under the action of the mean
field and  the average action of the collision term. For the fragmentation
process, however, correlations beyond the mean field are decisive whenever
the system enters an instability region. These higher order correlations
have been reintroduced as fluctuations in various ways: (1) by adding a
fluctuation term, leading to a Boltzmann--Langevin (BL) equation
\cite{ag90,rr90,abe96}, (2) by choosing the numerical 
fluctuation in a judicious
way by the number of test particles \cite{colonna}, or (3) by introducing
fluctuations directly into the phase space distribution \cite{macc}.
The detailed consequences of these approaches are presently under intense
investigation.

On the other hand, mean field dynamics governs the system as long as it
remains in stable regions of phase space. This is the case also after an
instability point, when the system has attained a new stable configuration.
In fact, within a mean field approach one should be able to determine 
when the system enters instability regions,
and also the characteristic thermodynamical state of this situation, i.e.
its density, pressure, temperature, isotopic ratios etc. It is also clear
that these critical values need not be the same in the finite colliding
system as in nuclear matter. Nuclear matter instability is  a pure volume
effect, while in a heavy ion collision finite size effects, in particular
surface effects, should be of great importance. Even though in a mean field
approach one will not be able to predict how the instable system breaks up,
one can therefore still determine the conditions at which this occurs. This
is not only an important input for models of statistical breakup or dynamical
fluctuations, it can also be compared to experimental determinations of the
fragmenting source.

The idea of the present work is then to investigate instability situations
within mean field dynamics, i.e. within transport calculations \cite{essl}. 
In doing so we want to take account of the fact that different 
thermodynamical conditions are expected to prevail in different 
regions and at different times in the collision process. We therefore 
determine the thermodynamic variables
locally. As a local criterion for instability we use a thermodynamical
condition, namely that the quantity 
$K_{{\mathrm eff}} = 9\frac{\partial P}{\partial\rho}$ becomes negative. 
We thus define $K_{{\mathrm eff}}$ 
as an effective compressibility for the finite systems which 
need not be the same as for infinite nuclear matter or for 
the ground state of nuclei. Other criteria
that have been used are the growth of numerical fluctuations \cite{colonna} 
or a positive Lyapunov exponent \cite{bonasera}. We believe that the
present criterion defines a dynamical instability. 

A particular difficulty is the determination of a temperature, and different
approaches have  been used for this. It is clear that at energies above about 
100 A.MeV the system is globally not in equilibrium and therefore a
temperature determined from the global momentum distribution is not
reasonable \cite{lbc91}. However, even the local momentum distribution need
not to be equilibrated, in fact, as will be seen later, the two subsystems
are still well separated in momentum space through much of the process. In
this situation the momentum distribution may still be represented as that of
two equilibrated subsystems of finite temperature \cite{pu94} 
and we also follow this approach here. It is still a question of debate, how
such temperatures should be compared to the various experimental
thermometers, as e.g. slope parameters of particle spectra
\cite{pion}, excited state or isotopic ratios \cite{poch95}. Certainly
slope parameters correspond least well to the temperatures determined
here.

As a realization of transport equations we use the relativistic
Landau--Vlasov (RLV) method, which was originally introduced in
Ref.\,\cite{Gregoire} and relativistically extended in
Ref.\,\cite{fu95}. It uses covariant Gaussian test particles in coordinate
and momentum space and thus allows to calculate truly local quantities at
every space--time point. It is thus particularly suited for the present
purpose.

We apply the investigation to Au on Au collisions at 600~A.MeV energy which
has been extensively investigated by the ALADIN \cite{poch95,traut97} 
and FOPI \cite{reisdorf97}  collaborations in particular 
with respect to possible phase transitions. We have previously 
investigated this reaction at 400~MeV in Ref.\,\cite{fuchs96} 
in particular with respect to the interactions to be used, 
the question of non--equilibrium effects, and the determination 
of the equation--of--state. Since here our main emphasis is on the 
investigation of instability conditions, we use a simpler standard 
interaction. We study a semi--central collision, where we can 
distinguish clearly a central  participant and a spectator region, 
which should behave very differently with respect to their 
thermodynamical evolution.

%%%%%%%%%%%%%%%%%%%%%%%%%%%%%%%%%%%%%%%%%%%%%%%%%%%%%%%%%%%%%%%%%%%%%%%%%
\section{Determination of local thermodynamic properties}
%%%%%%%%%%%%%%%%%%%%%%%%%%%%%%%%%%%%%%%%%%%%%%%%%%%%%%%%%%%%%%%%%%%%%%%%%
In equilibrated nuclear matter the pressure $P$ is isotropic and the 
energy-momentum tensor $T^{\mu\nu}$ takes the same form as in an ideal fluid
\cite{glw80}
\beq
T^{\mu\nu} (x) = \left[ \epsilon (x)-P(x)\right] u^\mu (x) u^\nu (x)
 - P(x) g^{\mu\nu}
\quad .
\label{tensor1}
\eeq
The streaming velocity $u_\mu = j_\mu / \rho_0$  is obtained from the
baryonic current
\beq
j_\mu (x) = 4 \int \frac{{\d}^4 k}{(2\pi)^3} k^*_\mu 
 f (x,k)
\label{current1}
\eeq
where the phase space distribution $f$  is related to the 
local momentum distribution $n$ as  
\beq
f(x,k) = n(x,{\vec k}) \delta \left( k^{* 2} -m^{* 2} \right) 
2 \Theta (k^{*}_0 ) 
\quad .
\label{phase}
\eeq 
The Lorentz invariant baryon rest density is defined as 
\beqa
\rho_0 = \sqrt{j_\mu j^\mu} = \rho_{\rm B} |_{\mathrm{rest frame}} 
\quad .
\label{density}
\eeqa
In the rest frame of equilibrated  nuclear matter the momentum 
distribution is given by a diffuse Fermi sphere 
or, in a general frame, by a diffuse Fermi ellipsoid
\beq
n(x,{\vec k},T) = 
\frac{1}{ 1+ \exp\left[ - (\mu^* - k_{\mu}^* u^\mu )/T \right]}
\label{fermi}
\eeq
with the temperature $T$, the effective chemical potential $\mu^* (T)$ 
and $ k_{0}^* = E^* = \sqrt{ {\vec k}^{*2} + m^{*2} }$. 
In the case of vanishing temperature Eq. (\ref{fermi}) 
reduces to a sharp Fermi ellipsoid 
\beq 
\lim_{T\to 0} n(x,{\vec k},T) = 
\Theta \left(E_{\rm F} - k_\mu^* u^\mu \right)
\label{fermi2} 
\eeq
with the chemical potential given by the Fermi energy 
$E_{\rm F} = \sqrt{m^{*2} + k_{{\rm F}}^2 }$. 

In these expressions the effective mass $m^* = M - g_\sigma \Phi (x)$ 
and the kinetic four-momenta $k^*_\mu = k_\mu - g_\omega \omega_\mu (x) 
-\frac{e}{2}(1+\tau_3 )A_\mu (x)$ are shifted by the scalar $\Phi$ 
and vector meson $\omega_{\mu}$  fields, respectively and a Coulomb 
four-vector potential $A_\mu$ is added and treated in a 
action-at-a-distance formulation \cite{fsc89}.

By definition Eq.(\ref{tensor1}) already contains the case of 
a collective motion of the matter as a whole, e.g., in a radially expanding
source $u_\mu = (\gamma,\gamma {\vec \beta}) $ is given by a radial 
flow $\vec\beta$ and only vanishes in the local rest frame. However, 
in the  time evolution of a heavy ion collision the situation of 
equilibrated nuclear matter, Eq.(\ref{tensor1}), is the exception. 
Through most of the reaction the 
system is {\em not} in local equilibrium.  Except for special 
cases as, e.g., in the final fireball or in the  spectator matter, the
pressure is not isotropic but rather has to be decomposed into 
contributions transversal and longitudinal to the relative velocity 
of the currents. We choose the $z$--axis as the beam direction. Then 
the pressure components follow from the energy momentum tensor 
($i=1,2,3$ correspond to $x,y,z$) as  
\beq
P_\perp = \frac{1}{2}\left( T^{11}+T^{22} \right) 
\quad , \quad 
P_\| = T^{33}
\quad .
\label{pressure}
\eeq   
$P_\perp $ and $P_\|$ are determined in the local rest frame 
which in colliding nuclear matter is the center-of-mass frame 
of the two currents where the total baryon 
current vanishes \cite{sehn96}. 
Thus, the difference of $P_\perp $ and $P_\|$ also 
yields a measure for the equilibration of the system.

 Although the total system is not in local equilibrium the single currents of
their own may be so. In actual calculations it is found, that
the phase space can be well approximated by configurations close to the 
initial one, i.e. by two separated Fermi spheres \cite{pu94} or, covariantly,
by two separated 
Fermi ellipsoids, which correspond to colliding nuclear matter currents
\cite{fuchs96}. We therefore approximate the phase space 
by colliding nuclear matter configurations  as 
described in Refs. \cite{fuchs96,sehn96}, however, in the 
present work of non-zero temperature. These configurations 
 are represented as \cite{sehn91}
\beq
n^{(12)} = n^{(1)} + n^{(2)} -\delta n^{(12)}
\quad , 
\label{conf12}
\eeq
where $\delta n^{(12)} = \sqrt{n^{(1)} \cdot n^{(2)} }$ is a Pauli 
correction which guarantees the validity of the 
Pauli principle in the case that the currents 
$n^{(1)}$ and $ n^{(2)}$ overlap. The 
single distributions $n^{(i)}$ are given by Eq.(\ref{fermi}). 
In principle the temperatures $T_1$ and $T_2$ can be different. The
collective 
parameters $u_{1\mu}$, $u_{2\mu}$, 
and 
$\mu^{*}_1 (T_1)$, $\mu^{*}_2(T_2)$  
are determined from the phase space distribution by a decomposition into 
contributions stemming from projectile and target 
$f(x,k) =f^{(1)}(x,k) +f^{(2)} (x,k) $ \cite{fuchs96}. 
In the limit of a vanishing relative velocity the ansatz of
eq.\,(\ref{conf12})
provides a smooth transition from the non-equilibrium colliding 
nuclear matter configuration to the equilibrated one,  
i.e. $ n^{(12)} \longmapsto n$. Furthermore the asymptotic state 
of two cold ($T=0$) currents is naturally included in this description.

The local temperature is obtained from a least square fit 
of the expression given by Eqs. (\ref{fermi}) and (\ref{conf12}) 
to the momentum distribution $n(x, {\vec k})$, Eq. (\ref{fit1}), 
obtained from the relativistic 
transport calculation \cite{fu95}. In this fit only the temperatures 
are treated as free parameters. The streaming velocities are 
determined directly from the respective currents, Eq. (\ref{current1}), 
and the chemical potentials $\mu^{*}_i(T_i) $ entering into Eq.
(\ref{conf12}) are obtained by the requirement of 
total baryon number conservation
\beq
j_0 (x) = 4 \int \frac{{\d}^3 k}{(2\pi)^3} n^{(12)} (x,{\vec k},T)
\eeq
by iteration. 

As discussed in the introduction we use as a realization of the transport
equation the relativistic Landau--Vlasov (RLV) method, 
which makes use of covariant Gaussian test particles in coordinate and
momentum space \cite{fu95}.
The RLV momentum space distribution  is given by 
\beq
n (x,{\vec k}) = \frac{(2\pi)^3}{4} \frac{1}{N(\pi \sigma \sigma_{\rm k})^3}
\sum_{i=1}^{A\cdot N }
\e^{[(x_\mu - x_{i\mu})^2 - ((x_\mu - x_{i\mu}) u_{i}^\mu )^2]/\sigma^2}
\e^{(k^{*2} - (k^{*}_\mu  u_{i}^\mu)^2)/\sigma_{\rm k}^2}
\label{fit1}
\eeq
and in our particular calculations we use
 $N=100$  testparticles per nucleon and $\sigma = 1.40$ fm, 
$\sigma_{\rm k} = 0.346$ fm$^{-1}$ as the width of the Gaussian in coordinate 
and momentum space, respectively.  This allows us to calculate all
quantities locally without recourse to discretization in cells. There is,
however, a technical point to be noted. The finite width of the momentum
space gaussians leads to a smearing of the local momentum distribution
$f(x,k)$, which would be interpreted as an artificial temperature, when
fitted with the expression of Eqs.\,(\ref{fermi}) or (\ref{conf12}).
Thus even for a initialized cold nucleus finite 
temperatures are obtained depending on $\sigma_{\rm k}$, the  width of 
the gaussians in momentum space (in our case about $T\sim5$ MeV). To take
this into account the configurations, 
Eqs. (\ref{fermi}) and (\ref{conf12}), are first folded 
over the momentum space gaussian. The folding procedure is 
defined covariantly as 
$ {\tilde n }(x,{\vec k},T) = \int {\d}^4 k'  m^* n (x,{\vec k}' ,T) 
g( k' - k) \delta ( k^{*' 2} - m^{* 2} ) 2\Theta ( k^{*'}_0 )  $. 
The momentum space gaussian $g(k')$ is defined as in (\ref{fit1}), 
however, with the streaming velocity of the current, Eq.\,(\ref{current1}), 
instead of the particle velocity. 
The folding is most naturally performed in the 
rest frame of the current where the expression reduces to a 
folding over a standard gaussian, i.e.
\begin{displaymath} 
 {\tilde n }(x,{\vec k},T) = \int 
\frac{ {\d}y^3 k' }{\sqrt{\pi \sigma_{{\rm k}}^2}^3}\frac{m^*}{E^{*'}} 
n (x,{\vec k}' ,T) 
\e^{ - ( {\vec k}^{*'} - {\vec k}^{*})^2 /\sigma_{\rm k}^2 }
\quad . 
\end{displaymath}
In the case of colliding nuclear matter configurations 
(\ref{conf12}) this procedure has to be performed separately 
for each current. Thus artificial temperature effects 
are eliminated and  reliable initial values 
are obtained ($T\sim 0.5$ MeV) which are a very good 
starting point for the numerical analysis. 
%%%%%%%%%%%%%%%%%%%%%%%%%%%%%%%%%%%%%%%%%%%%%%%%%%%%%%%%%%%%%%%%%%%%%%%%%
\section{Thermodynamic properties of Au on Au at 600 A.MeV}
%%%%%%%%%%%%%%%%%%%%%%%%%%%%%%%%%%%%%%%%%%%%%%%%%%%%%%%%%%%%%%%%%%%%%%%%%
The analysis described above is applied to a typical 
intermediate energy reaction, i.e. a semi-central (b=4.5 fm) 
Au on Au reaction at 600 A.MeV.
This same reaction was also investigated in Ref.\,\cite{fuchs96} at
400~A.MeV with the RLV method \cite{fu95}. It
has been extensively studied by the FOPI and ALADIN collaborations at GSI 
\cite{poch95,traut97,reisdorf97}. In relativistic 
models the mean field originates from the cancelation of 
large scalar and vector fields. Realistic mean fields  derived from the 
Dirac-Brueckner $G$-matrix and which also account for 
non-equilibrium aspect of the phase space \cite{sehn96} 
were, e.g. used in the calculations in Ref. \cite{fuchs96}. 
However, for simplicity, in the present work we use the 
standard parameterization of the non-linear Walecka model (NL2) 
\cite{lbc91}, which was also used for comparison in Ref.\,\cite{fuchs96}.

In Fig. 1 we demonstrate the basic idea of the present approach. 
The left hand column gives density contours at different times of the
collision (10,50,60~fm/c). The two columns to the right give contours of
local momentum distributions, the one obtained in the RLV calculation in the
middle and the one fitted according to eq.\,(\ref{conf12}) to the right with
the fit values of the temperature also given. The two upper rows show the
momentum distribution in the center, the lower one in the spectator moving to
the left.

It is seen that after 10 fm/c the 
temperature is already high in the center (T=34.6 MeV), 
however, the ellipsoids 
corresponding to projectile and target are still well separated. Thus 
the configuration is highly anisotropic and is well represented by two hot
and counterstreaming currents of nuclear matter. At 
50 fm/c the system has reached local equilibrium in the center and has also 
strongly cooled down (T=5.4 MeV). Thus we find that as long as 
temperatures are high non-equilibrium aspects of the phase 
space are of major importance. At the later stages 
where equilibrium is reached the system cannot really be considered as a 
" fireball " since the temperature is already low. However, 
in both cases the phase space is  
well approximated by the parameterization of eq.\,(\ref{conf12}) which
describes one or two equilibrated subsystems. 
The spectator is clearly identified in the density contour plots at the
later stages. In the spectator the momentum space is well represented by one
Fermi fluid and the temperature is well defined and found to be low.

We next look at the  evolution of density and temperature in the central
and spectator regions. We will put more emphasis on the discussion of the
spectator since it will be seen to be the more interesting part. In 
the following analysis we identify the spectator region at each 
time step from inspection of the evolution of the density profile 
and extract the respective 
observables at the position of maximum density.

In Fig.\,2 the evolution of the density is shown for the spectator and in the
center (insert). In the center we see the typical compression--decompression
behavior. The total density rises to about 2.5~$\rho_{\mathrm{sat}}$ 
and then continuously falls to zero. The spectator density  
decreases in the decompression phase from about saturation density 
($\rho_{\mathrm{sat}} = 0.145$ fm$^{-3}$ in the present 
model) to a value of $\sim \frac{1}{3} \rho_{\mathrm{sat}}$ 
where it then stays relatively constant over a period 
of 30 fm/c before the spectator completely evaporates. 

In Fig.\,3 we show the corresponding evolution of the local temperature again
for spectator and central (insert) regions. 
For the temperature  in the central " fireball " 
region we obtain a maximum value of about 40 MeV and a 
mean temperature of about 30 MeV over the duration of the 
compression phase. Thus the present phase space analysis results 
in significantly smaller temperatures than those extracted from 
particle spectra 
\cite{pion}. This indicates that temperatures determined from the 
slope parameters of such spectra   apparently 
overestimate the real temperature by at least a factor of two. 
Similar results have been found in Ref. \cite{ksg95}. 
However, the present results are in 
qualitative agreement with temperatures extracted experimentally 
from a flow analysis within the blast scenario. E.g., in Ref. 
\cite{reisdorf97} a value of $T=36.7 \pm7.5$ MeV has been found 
in  Au on Au collisions at 400 A.MeV.  

The temperature in the spectator shows a rather different behavior.
When the spectator region is clearly developed in the transport 
calculation its temperature is about 15 MeV and then continously 
decreases to a value around 5 MeV where it stays fairly stable. 
Thus density and temperature in the spectator both show a sort of plateau
between about 50 and 85~fm/c. It is seen later that this phase seems to
correspond to a region of instability.

To study this further we now investigate $P-\rho_0$ diagrams, which show the
thermodynamical evolution of the system as a trajectory with time as a
parameter. Since we have defined 
$K_{{\mathrm eff}} = 9\frac{\partial P}{\partial\rho_0 }$
as the effective compressibility a negative slope of this trajectory
signifies that the system enters a region of instability.
 At this point the system is expected to be sensitive to fluctuations and 
eventually to form fragments.

In Fig. 4 we show the evolution of 
the central region in the $P-\rho_0 $  diagram. 
In the following $\rho_0$ corresponds to the local rest density, 
Eq. (\ref{density}) and thus a distortion of the results by 
Lorentz effects is eliminated. It can be seen that the pressure is mostly
positive and
is maximal in a very early stage of the reaction ($t=10-15$ fm/c) 
and then rapidly drops down to values around zero ($ t\sim 10-15$ fm/c). 
In the compression phase which lasts from about 10--30 fm/c (see also 
Fig. 2) the pressure is highly anisotropic, i.e. the longitudinal 
component is more than twice as large as the transversal one. After 
30 fm/c both components have approached values close to zero. 
The respective maximum values 
$P_{\|} = 80 $ MeV fm$^{-3}$, $P_{\perp} = 30 $ MeV fm$^{-3}$ are in 
qualitative agreement with the analysis of Ref. \cite{lbc91} where, 
however, only global quantities have been considered. 
 A  thermodynamic instabilitity region, i.e. a negative slope of 
 the $P-\rho_0$ trajectory, occurs after 60 fm/c. This is not seen within the scale of the figure, because the density  is already very low (Fig. 2). Therefore one does not expect the formation of larger fragments in the central region.

In Fig. 5 we show the corresponding $P-\rho_0$ diagram 
for the spectator matter. The 
pressure is now negative for densities below saturation which reflects 
the van-der-Waals like behavior of the nuclear matter equation of 
state also in the finite system. Furthermore, 
the spectator is not completely equilibrated 
even at the later stages of the reaction 
since longitudinal and transversal components are still significantly 
different in magnitude. They show, however, a very similar 
behavior in their phase trajectories. It is seen that after 45 to 50 fm/c 
both the longitudinal and the transverse pressures increase 
with decreasing density and thus the 
compressibility becomes negative. 
As discussed in connection with Fig.\,3 the temperature is rather stable
after this point.
This ensures that a thermodynamic compressibility 
$K_{{\mathrm eff}} = 9 
\frac{\partial P}{\partial\rho_0} |_{T={\mathrm const} }$ 
can be defined in a meaningful way, i.e. the change of the local 
pressure occurs in good approximation at constant temperature. 
The system at this stage therefore enters an instability region and 
should break up into fragments. The break up density lies 
between $(\frac{1}{3} - \half) \rho_{\mathrm{sat}}$. 
In the present analysis the value of $T=5$ MeV 
corresponds to the break up temperature. Thus 
it is in rather good agreement with the experimental value of the critical 
temperature at the liquid-gas phase transition measured by the 
ALADIN Collaboration for the same reaction \cite{poch95}.

As discussed, a description of the break up process 
is beyond the scope of a mean field approach. 
 In the mean field calculation the 
system remains in a relatively stable 
configuration at densities around $\frac{1}{3} \rho_{\mathrm{sat}}$ 
which lasts over a period from 60 to 85 fm/c before the spectator 
completely evaporates. 

In fig.\,5 we have underlaid the isothermal equation--of--state for
thermalized nuclear matter for temperatures of 5 and 9~MeV, which correspond
to the range of temperatures determined for the spectator (see Fig.\,3). The
nuclear matter spinodal region is that part of the curves, where the slope is
negative. As stressed before the instability conditions, as determined here,
do not have to be identical to those of nuclear matter. However, it is seen
that in the final stages of its evolution (after about 65~mf/c) the spectator
rather closely follows the nuclear matter behavior, as one would perhaps
expect. Before that, the thermodynamic conditions are appreciably different
from those of thermalized nuclear matter. In the early stages of the
spectator ($t < 40$~fm/c) the pressures are higher because there is not yet a
clear separation of spectator and participant. While the critical density,
where instability sets in, is about the same the pressures are different and
there is still substantial anisotropy.
Therefore it is to be expected that the fragmentation process which is
initiated by fluctuations at this point is strongly influenced by these
conditions. Thus for the treatment of multi--fragmentation the treatment of
these dynamical instabilities is very important.

%%%%%%%%%%%%%%%%%%%%%%%%%%%%%%%%%%%%%%%%%%%%%%%%%%%%%%%%%%%%%%%%%%%%%%%%%
\section{Conclusions}
%%%%%%%%%%%%%%%%%%%%%%%%%%%%%%%%%%%%%%%%%%%%%%%%%%%%%%%%%%%%%%%%%%%%%%%%%
The purpose of the present work is to investigate in the framework of
transport theories the thermodynamical state of matter in a heavy ion
collision, in particular with respect to the occurrence of instabilities. It is argued that this can be answered in a mean field treatment, while the
further evolution and possible fragmentation depends on fluctuations and is
beyond this approach. To this end we determine thermodynamic variables
directly from the local momentum distribution. In particular, temperature is
obtained by a fit to two hot Fermi distributions, respecting the Pauli
principle, thus taking into account the most typical non--equilibrium effect
in a heavy ion collision. In this way the three intensive variables T, P,
$\rho$ are determined independently from each other, while in a specific
system, e.g. in nuclear matter, they are, of course, constrained by the
equation--of--state. A comparison therefore shows the effect of finite size
and non--equilibrium effects on the thermodynamical state. As a criterion for
instability we use a negative effective compressibility defined from the
local thermodynamic variables.
Thus we use the concept of a dynamical instability which is different, in
principle, from the static spinodal instability in nuclear matter.

We applied these methods to the typical, semi-central, well--studied, 
intermediate energy reaction Au+Au at 600~A.MeV and investigate the central
and the spectator zones. We determine thermodynamical variables which are
largely consistent with experimental determinations. In particular we see an
instability develop in the heated spectator, which should then lead to
spectator multi--fragmentation as observed in the ALADIN collaboration. Also
the breakup temperatures and densities are in reasonable agreement. We see
that the thermodynamical variables thus determined are different from the
nuclear matter equation--of--state. The determination of instability situations  is important
for the description of multi--fragmentation in other theories: for the
application of statistical approaches one has to know whether the breakup
configuration is equilibrated and what are its parameters; for dynamical
treatments of fluctuations, as in the various approaches to the BL--equation,
it is important to know, at what points in the evolution fluctuations are
important and what is their magnitude. The present analysis will be extended
in the future to a more systematic study of heavy ion collisions, in
particular also to less symmetric points. This will allow to also study
collective flow and temperature in the context of radial flow scenarios.

%%%%%%%%%%%%%%%%%%%%%%%%%%%%%%%%%%%%%%%%%%%%%%%%%%%%%%%%%%%%%%%%%%%%%%%%%
%                                                                       %
%   END OF TEXT                                                         %
%                                                                       %
%%%%%%%%%%%%%%%%%%%%%%%%%%%%%%%%%%%%%%%%%%%%%%%%%%%%%%%%%%%%%%%%%%%%%%%%%
\newpage
%%%%%%%%%%%%%%%%%%%%%%%%%%%%%%%%%%%%%%%%%%%%%%%%%%%%%%%%%%%%%%%%%%%%%%%%%
%                                                                       %
%   BEGIN OF THEBILIOGRAPHY                                             %
%                                                                       %
%%%%%%%%%%%%%%%%%%%%%%%%%%%%%%%%%%%%%%%%%%%%%%%%%%%%%%%%%%%%%%%%%%%%%%%%%

%%%%%%%%%%%%%%%%%%%%%%%%%%%%%%%%%%%%%%%%%%%%%%%%%%%%%%%%%%%%%%%%%%%%%%%%%
%                                                                       %
%   END OF THEBILIOGRAPHY                                               %
%                                                                       %
%%%%%%%%%%%%%%%%%%%%%%%%%%%%%%%%%%%%%%%%%%%%%%%%%%%%%%%%%%%%%%%%%%%%%%%%%

\begin{figure}[b]  %%Fig.1%%%%%%%%%%%
\begin{center}
\leavevmode
\epsfxsize = 13cm
\epsffile[38 23 551 430]{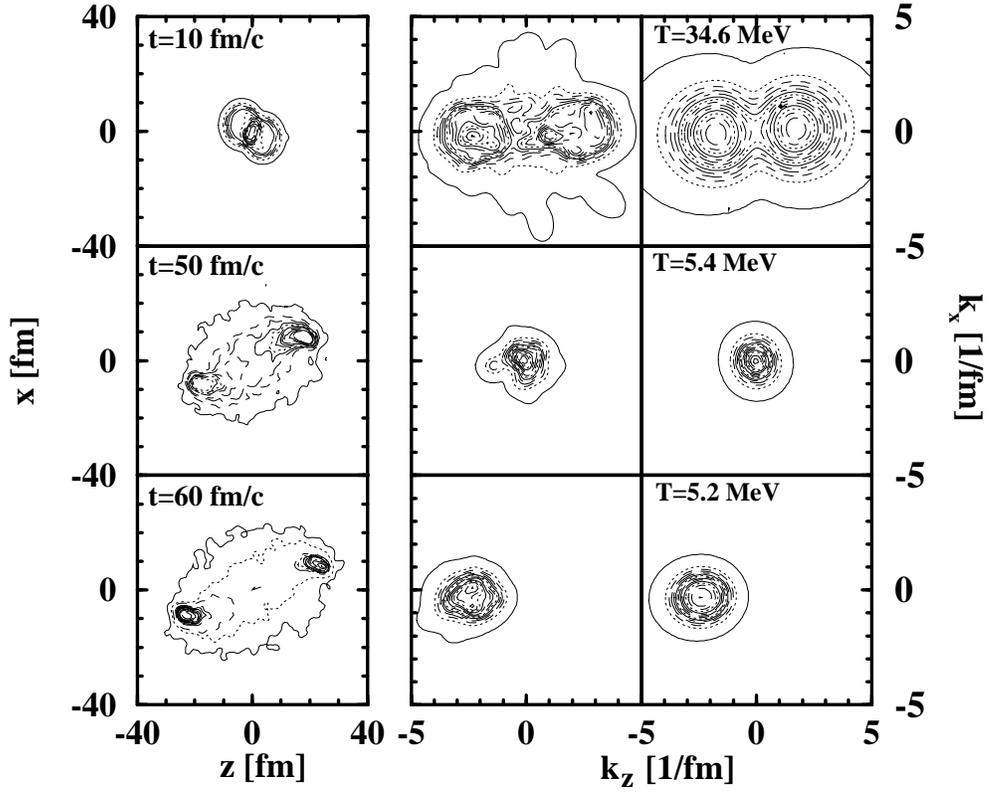}
\end{center}
\caption{    %%neu%%
 Phase space distributions in the semi--central (b = 4.5~fm) Au+Au reaction
at 600~A.MeV. The three rows correspond to times t = 10, 50, and 60~fm/c.
The left-most column gives the density contours, the two right hand ones 
local momentum distributions as obtained from the transport calculations
(middle) and by a fit with one or two hot Fermi--ellipsoids (right) with temperatures indicated. In the first two rows the momentum distributions are  shown for the center, in the last row for the spectator.
}
\end{figure}
%%%%%%%%%%%%%%%%%%%%%%%%%%%%%%%%%%%%%%%%%%%%%%%%%%%%%%%%%%%%%%%%%%%%%%%%%
\begin{figure}[h]   %%%Fig.2%%
\begin{center}
\leavevmode
\epsfxsize = 12cm
\epsffile[60 120 450 590]{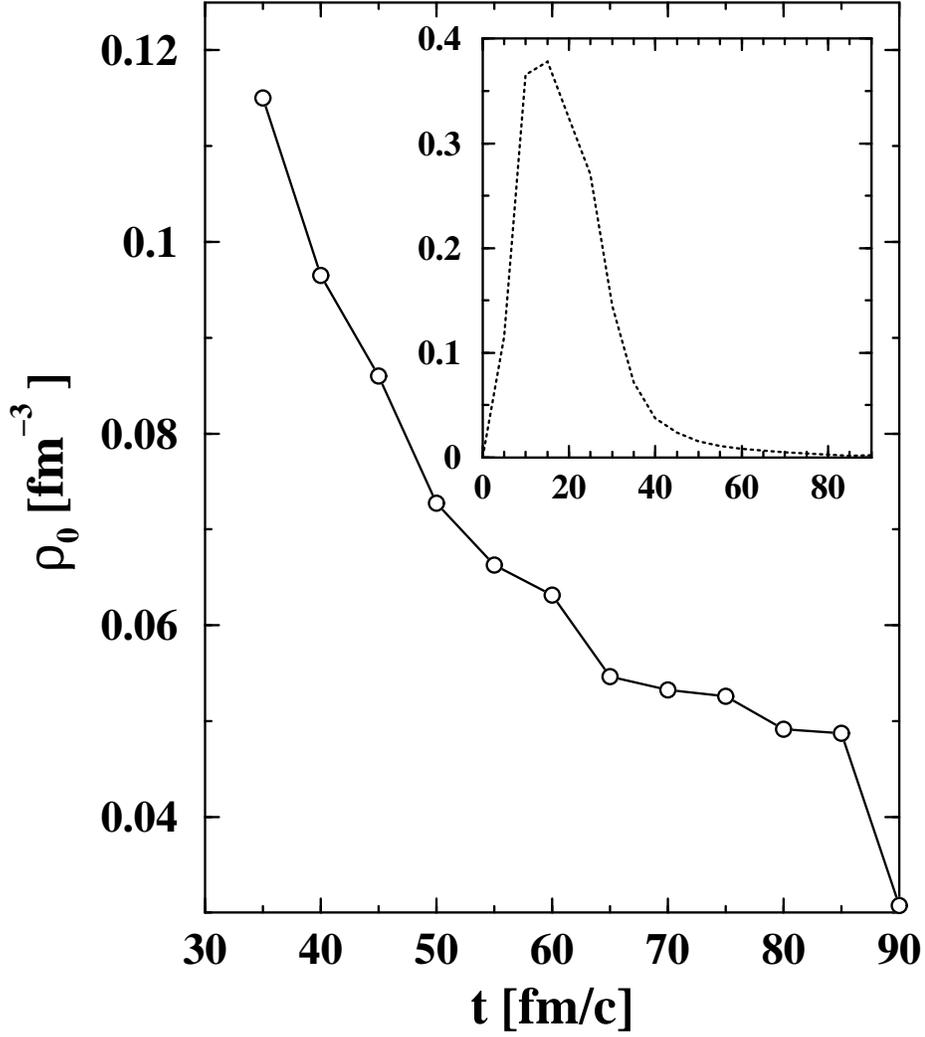}
\end{center}
\caption{
Time evolution of the local total density in the spectator matter in a 
semi-central Au on Au rection at 600 A.MeV. The insert  
shows the corresponding density in the center.
}
\end{figure}
%%%%%%%%%%%%%%%%%%%%%%%%%%%%%%%%%%%%%%%%%%%%%%%%%%%%%%%%%%%%%%%%%%%%%%%%%
\begin{figure}[h]    %%Fig.3%%%%%%%%%
\begin{center}
\leavevmode
\epsfxsize = 12cm
\epsffile[60 120 450 590]{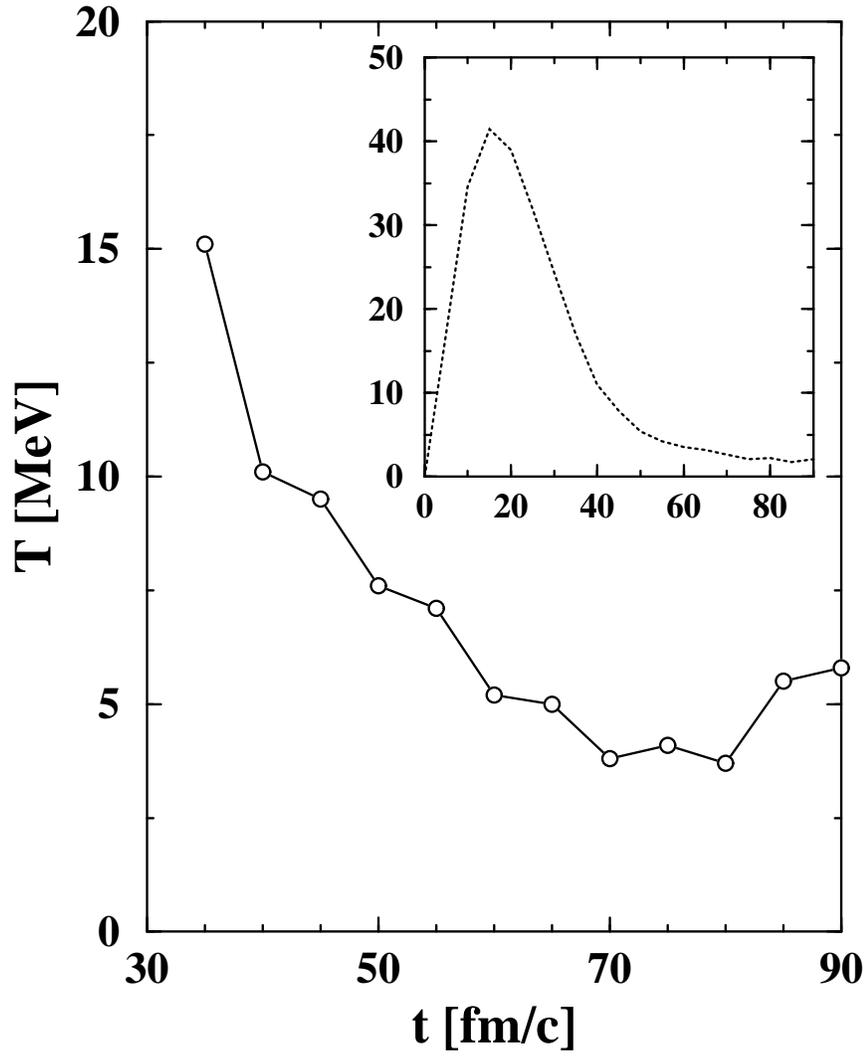}
\end{center}
\caption{
Time evolution of the local temperature of the spectator matter in a 
semi-central Au on Au rection at 600 A.MeV. The insert  
shows the corresponding temperature obtained in the center.
}
\end{figure}
%%%%%%%%%%%%%%%%%%%%%%%%%%%%%%%%%%%%%%%%%%%%%%%%%%%%%%%%%%%%%%%%%%%%%%%%%
\begin{figure}[h]    %%Fig.4%%%%%%%%%%%
\begin{center}
\leavevmode
\epsfxsize = 12cm
\epsffile[46 57 470 423]{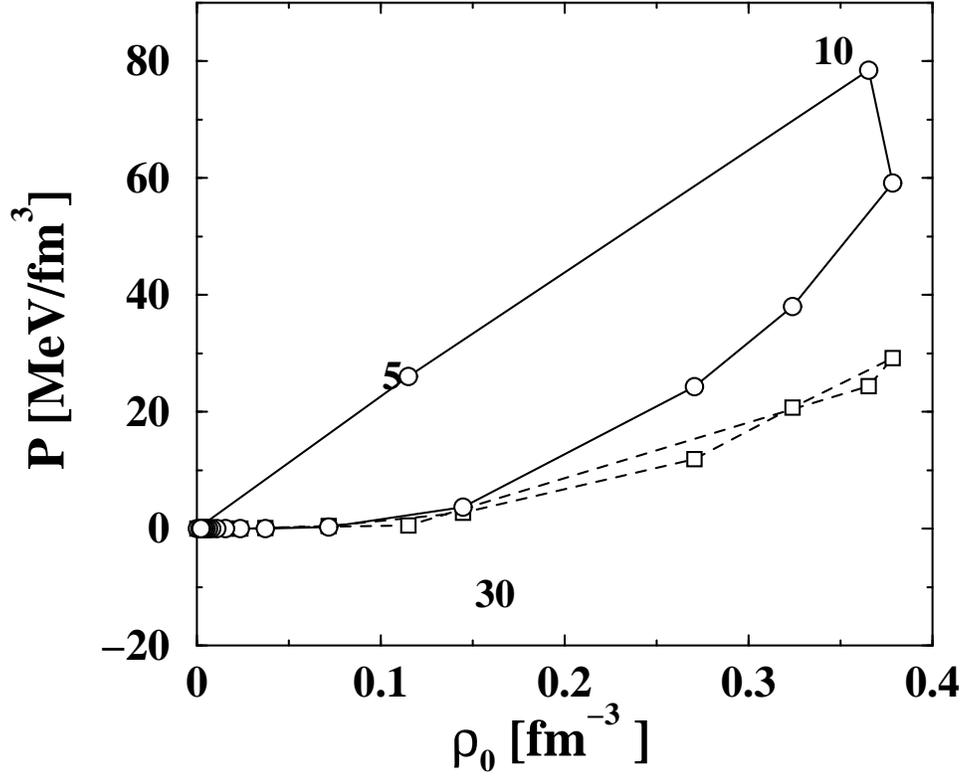}
\end{center}
\caption{
Density-pressure trajectory in the center obtained in a semi-central Au on Au reaction at 600 A.MeV. The evolution of longitudinal (solid line) 
and transverse (dashed line) pressure is shown separately. The corresponding times (fm/c) are indicated for some cases points (the difference between points is 5~fm/c). 
}
\end{figure}
%%%%%%%%%%%%%%%%%%%%%%%%%%%%%%%%%%%%%%%%%%%%%%%%%%%%%%%%%%%%%%%%%%%%%%%%%
\begin{figure}[h]    %%Fig.5%%%%%%%%%%%%
\begin{center}
\leavevmode
\epsfxsize = 12cm
\epsffile[46 57 470 423]{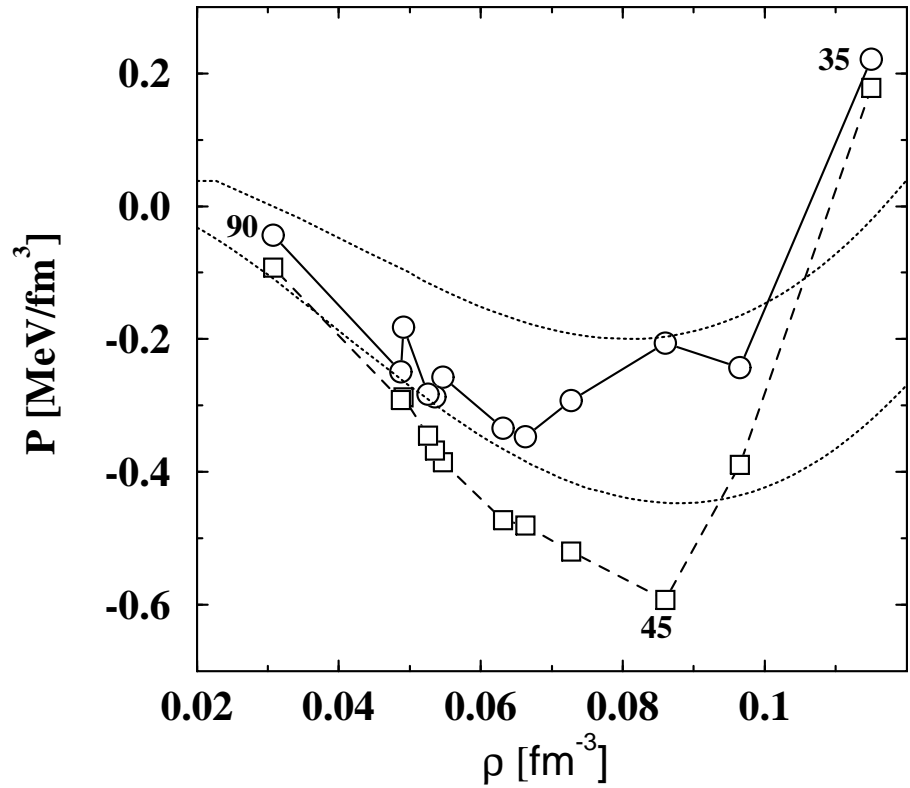}
\end{center}
\caption{     
Density-pressure trajectory for the spectator matter in a semi-central 
Au on Au reaction at 600 A.MeV as in Fig.\,4. The dotted curves
are the nuclear matter equation of state for T = 5 and 9~MeV (lower and 
upper curve, respectively).
}
\end{figure}
%%%%%%%%%%%%%%%%%%%%%%%%%%%%%%%%%%%%%%%%%%%%%%%%%%%%%%%%%%%%%%%%%%%%%%%%%
%                                                                       %
%   END OF DOCUMENT                                                     %
%                                                                       %
%%%%%%%%%%%%%%%%%%%%%%%%%%%%%%%%%%%%%%%%%%%%%%%%%%%%%%%%%%%%%%%%%%%%%%%%%
\end{document}